\documentclass[a4paper]{jpconf}
\usepackage{graphicx}
\begin{document}
\title{On conservative models \\ of ``the pair-production anomaly'' \\ in blazar spectra at Very High Energies}

\author{T.A. Dzhatdoev}

\address{Skobeltsyn Institute of Nuclear Physics, Lomonosov Moscow State University, Leninskie gory 1-2, 119991 Moscow, Russia}

\ead{timur1606@gmail.com}

\begin{abstract}
For some blazars, the gamma-ray absorption features due to pair-production on the Extragalactic Background Light (EBL) are fainter than expected. The present work reviews the main models that could explain this paradox, with emphasis on conservative ones, that do not include any new physics. The models that are intrinsic to the source, do allow a very hard primary spectrum, but fail to explain a regular redshift dependence of the anomaly starting energy. The model that includes a contribution from secondary photons produced by cosmic rays (CR) near the Earth seems to require a well collimated CR beam, what is hard to achieve. Finally, the model with secondary photons produced in electromagnetic (EM) cascades initiated by primary gamma-rays is considered. In principle, it allows to decrease the statistical significance of the anomaly and, while requiring quite low EGMF strength $B$, does not contradict to most contemporary constraints on the $B$ value. Additionally, it is shown that the recently observed correlation between directions to hard gamma-ray sources and voids in the Large Scale Structure is a natural feature of the EM cascade model.
\end{abstract}

\section{Introduction}

The number of detected extragalactic gamma-ray sources has greatly increased during the last decade. Blazars --- gamma-ray loud active galactic nuclei (AGN)--- constitute the great majority of these sources. By the start of 2015, $\sim$50 blazars were discovered by ground-based detectors \cite{tev15} (these instruments typically work at very high energy (VHE) range $E>$100 $GeV$), and $\sim10^{3}$ blazars were observed by the Fermi LAT gamma-ray space telescope at $E$= 100 $MeV$--100 $GeV$ \cite{ack11}. Besides investigation of AGN intrinsic properties, these observations allow to study extragalactic gamma-ray propagation as well. High energy photons are subject to the $\gamma\gamma \rightarrow e^{+}e^{-}$ process on the Extragalactic Background Light (EBL) photons \cite{nik62}--\cite{gou67} with subsequent formation of electromagnetic (EM) cascades. Thus, primary gamma-rays with $E>E_{abs}$ ($E_{abs}= E(\tau_{\gamma\gamma} = 1)$; $\tau_{\gamma\gamma}$ is the optical depth of the pair-production process) are effectively ``absorbed'', and their energy is transferred to electrons, positrons (in what follows called simply ``electrons'') and photons with comparatively low energy. Absorption features of the $\gamma\gamma \rightarrow e^{+}e^{-}$ process were already observed by the Fermi LAT instrument \cite{ack12}, as well as the H. E. S. S. Cherenkov telescope \cite{abr13} with high statistical significance: $\sim$6$\sigma$ and $8.8\sigma$, respectively.

However, it appears that for some blazars modification of the gamma-ray spectrum in the optically thick regime has a somewhat anomalous character \cite{hor12}. By extrapolating the spectrum from the optically thin regime, \cite{hor12} found that the distribution of the flux points scatter around the predicted intensity is different for the 1$<\tau_{\gamma\gamma}<$2 and $\tau_{\gamma\gamma}>$2 regions, and a strong indication (significance $Z_{a}= 4.2\sigma$) that the observed absorption is smaller than the assumed one was obtained. This effect, observed at VHE range, was called ``the pair-production anomaly'' by \cite{hor12}.

In fact, an anomaly of such kind is a long-standing problem of VHE gamma-astronomy \cite{aha99}--\cite{pro00}; a number of solutions was proposed, including quite exotic ones, such as violation of Lorentz invariance (LIV) \cite{kif99}--\cite{ame00},\cite{pro00}, and oscillations of photons to Axion Like Particles (ALPs) (e.g., \cite{sik83}--\cite{san09}). A search for new physical phenomena of such kind would require that all possible background processes are well known and appropriately accounted for. One such process is emission of secondary photons in EM cascades. On a subsample of spectra analysed in \cite{hor12}, it was recently shown that inclusion of this secondary component to the spectral fitting procedure indeed allows to considerably decrease the statistical significance of the anomaly \cite{dzh15}. In the present work the cascade model of the VHE anomaly in blazar spectra is compared with other models of the anomaly. The paper is organized as follows: section 2 contains a brief overview of possible processes that might relax the anomaly; the EM cascade (EMC) model is described in section 3; section 4 reviews some contemporary constraints on the Extragalactic Magnetic Field (EGMF) strength. Finally, some conclusions are presented in section 5.

\section{The models of the VHE anomaly}

\subsection{EBL intensity models}

The simplest assumption that could relax the anomaly would be that the EBL energy density is overestimated. If this is the case, the intrinsic spectrum would experience weaker modification than is believed, and for sufficiently low EBL density the VHE anomaly would be relaxed (note, however, that the dependence of the anomaly significance on the EBL intensity may have a non-trivial behaviour \cite{hor13}).

We are mostly interested in the wavelength region  $\lambda$=1--10 $\mu m$ that roughly corresponds to 1--10 $TeV$ minimal energy of high-energy photon that is subject to the $\gamma\gamma \rightarrow e^{+}e^{-}$ process. Many EBL models exist, including \cite{kne10}--\cite{gil12}; in the range of $\lambda$=1--10 $\mu m$ the model of \cite{kne10}, that was utilized in \cite{hor12}, predicts one of the lowest EBL intensity among \cite{kne10}--\cite{gil12}. The \cite{kne10} model is, in fact, a sort of lower limit on the EBL intensity. A well justified conclusion that such a limit must be revised would likely require a large amount of dedicated observational work, therefore we move to other possible explanations of the anomaly. 

\subsection{A list of effects that could relax the anomaly}

Besides the EBL properties, a number of factors exist that may have some relation to the nature of the VHE anomaly. A non-exhaustive list of them is given below. \\
I. Effects, connected to the source.\\
I.1. Intrinsic spectrum may have a pile-up resulting from inverse Compton (IC) scattering on an ultrarelativistic outflow, thus increasing observed intensity at high energies \cite{aha02}. \\
I.2. Internal absorption of primary photons on radiation field near the source, under certain assumptions, again, may produce a very hard intrinsic spectrum \cite{aha08} with a pile-up. \\
I.3. Another known mechanism of a hard spectrum formation is synchrotron radiation of electrons produced by ultrahigh energy (UHE) protons near the source \cite{oik14}. \\
I.4. It is believed that blazars have well collimated jets pointed towards the observer (the scheme of a typical blazar geometry taken from \cite{pad97} is drawn in figure 1). Magnetic field in a typical jet likely has some turbulent component, that may scatter charged energetic particles (electrons and hadrons). This scattering angle is usually larger for charged particles with comparatively low energy, that emit predominantly low-energy photons. Thus, the beaming factor for high-energy photons would be larger, again increasing observed intensity at high energies. To the author's knowledge, this effect was never discussed in connection to the VHE anomaly yet.

\begin{figure}[h]
\includegraphics[width=12pc]{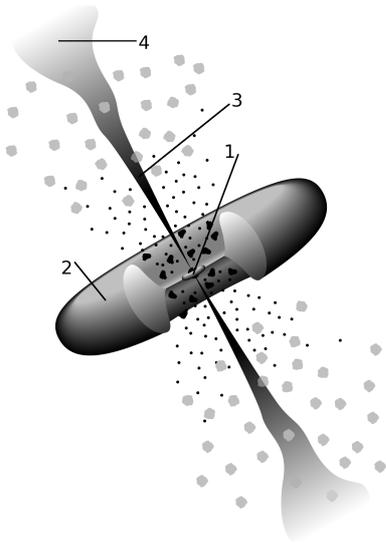}\hspace{2pc}%
\begin{minipage}[b]{24pc}\caption{\label{label} A sketch of a typical AGN geometry \cite{pad97}. 1 --- central compact object (black hole with accretion disk), 2 --- dusty torus around the central object, 3 --- well collimated relativistic jet with regular magnetic field $\vec{B}_{Jet}$ that can have some turbulent component $\delta\vec{B}_{Jet}$, 4 --- a ``lobe'' at the end of the jet that likely contains a highly turbulent magnetic field region.}
\end{minipage}
\end{figure}

However, it was observed that the energy at which the anomaly shows up, exhibits strong, regular dependence on redshift (this energy is defined by the relation $E_{a} \approx E(\tau_{\gamma\gamma}=2$) \cite{hor12}). The effects, intrinsic to the source, are not expected to display such a dependence, thus they can not constitute the only physical reason of the anomaly. \\
II. Propagation effects. \\
II.1. LIV effects, in principle, may suppress the $\gamma\gamma \rightarrow e^{+}e^{-}$ process \cite{kif99}. This mechanism is expected to operate above a certain fixed energy, and so it is disfavoured for the same reason as the source-induced effects (see discussion in \cite{hor12}). \\
II.2. $\gamma$-ALP oscillations (and vice versa) in magnetic field is another mechanism that makes the primary gamma-rays able to avoid strong absorption \cite{mir07}. Even if we assume the existence of ALP and that the $\gamma$-ALP oscillation process is indeed allowed, there is another potential difficulty: the efficiency of direct ($\gamma \rightarrow ALP$) and reverse ($ALP \rightarrow \gamma$) conversion is required to be high enough for many sources, independently of direction to a source, and, therefore, of the Galactic magnetic field parameters in this direction, as well as of the source' magnetic field. The recent data didn't allow the authors of \cite{wou14} to put any constraints on the ALP mechanism, but the same work shows that such constraints are obtainable in the future. \\
II.3. If blazars accelerate hadrons as well as electrons, secondary gamma-rays produced by the former may contribute to observed spectrum and thus relax the anomaly \cite{ury98}--\cite{ess10}. This model is discussed below. \\
II.4 As was already mentioned, the most basic process that may influence the shape of observed spectrum is emission of secondary photons in electromagnetic cascades \cite{aha99}, \cite{aha02}, \cite{ave07}, \cite{dzh15}. This process is discussed in section 3.

\subsection{The CR beam model}

Secondary photons produced by cosmic ray beam (by means of pair-production and photohadronic processes with subsequent development of electromagnetic cascades) relatively near to the Earth may enhance the intensity in the optically thick regime. Secondary gamma-ray spectra, typical for this mechanism, were studied in \cite{ess10b}--\cite{ess14}. The case of a very distant source (with redshift $z\sim$1) was considered in \cite{aha13}. Time structure of VHE gamma-ray signal was discussed in \cite{pro12}. CR beam may also produce an observable amount of neutrinos \cite{ess10b}--\cite{ess11},\cite{kal13}.

This model of the VHE anomaly has many attractive features: it allows to explain the observed spectrum without invoking any exotic processes such as $\gamma$-ALP oscillations. Moreover, the shape of the secondary photon component, assuming sufficiently low strength of the EGMF $B<10^{-14}$ $G$, depends mainly on the overall normalization of the primary CR spectrum. However, the model is not out of difficulties. As an example, let us consider the 1ES 1101-232 blazar spectrum (redshift $z$=0.186). Figure 2 shows a spectral energy distribution (SED) of the source measured by the H. E. S. S. Cherenkov telescope (red circles) together with uncertainties (dashed red lines); the assumed intrinsic spectrum is shown by black curve, the fit to the observed spectrum is shown by green line (for detals of calculations, that were performed with the publicly-available code ELMAG 2.02 \cite{kac12} and the EBL model of \cite{kne10}, see \cite{dzh15}). The intrinsic spectrum curve here is normalized to the absorbed one at $E$= 200 $GeV$, where absorption effects are already small. The fit here doesn't include any secondary photon component.

\begin{figure}[h]
\begin{minipage}{18pc}
\includegraphics[width=18pc]{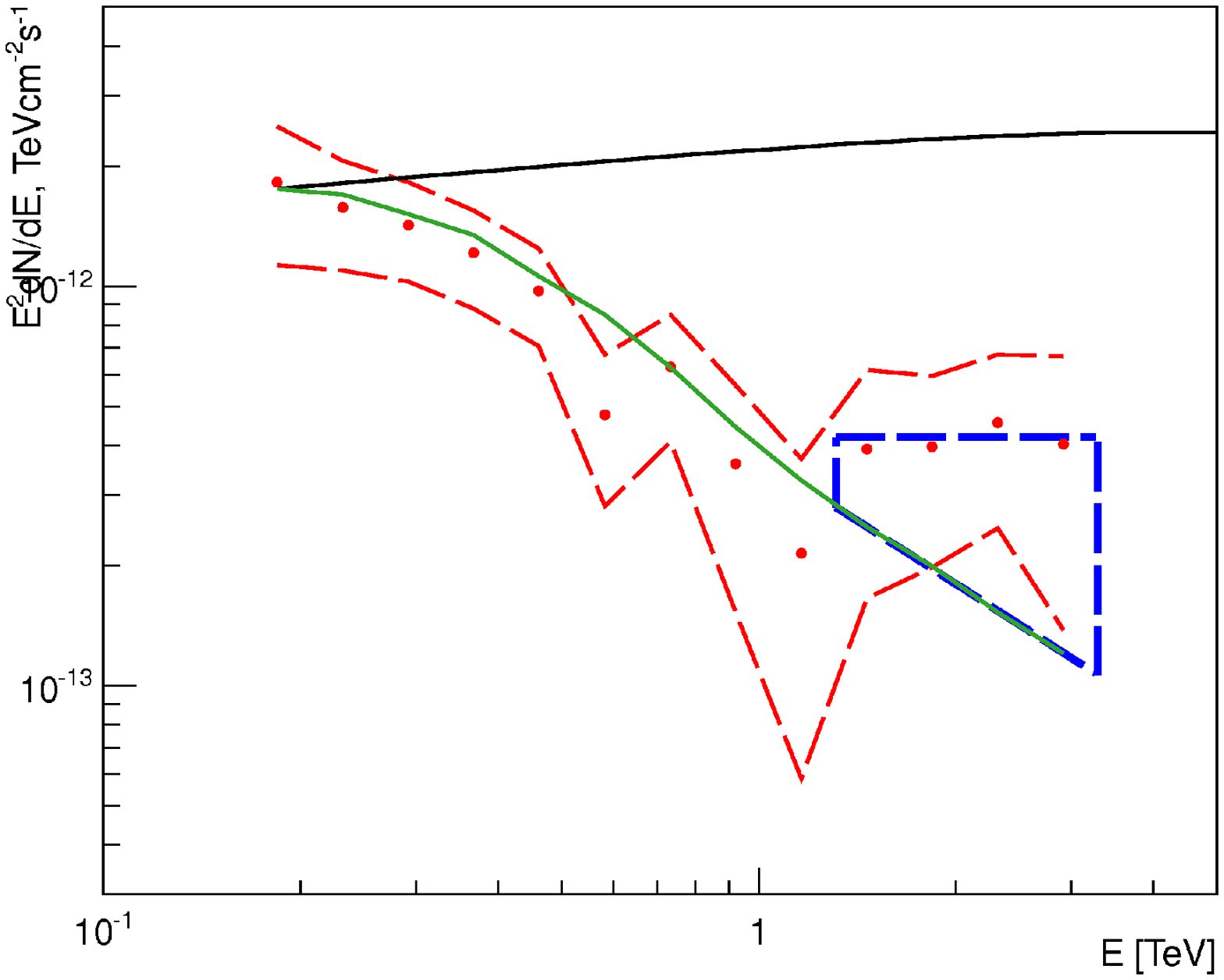}
\caption{\label{label} The fit to the 1ES 1101-232 SED \cite{aha06} without account of any secondary emission \cite{dzh15}.}
\end{minipage}\hspace{2pc}%
\begin{minipage}{18pc}
\includegraphics[width=18pc]{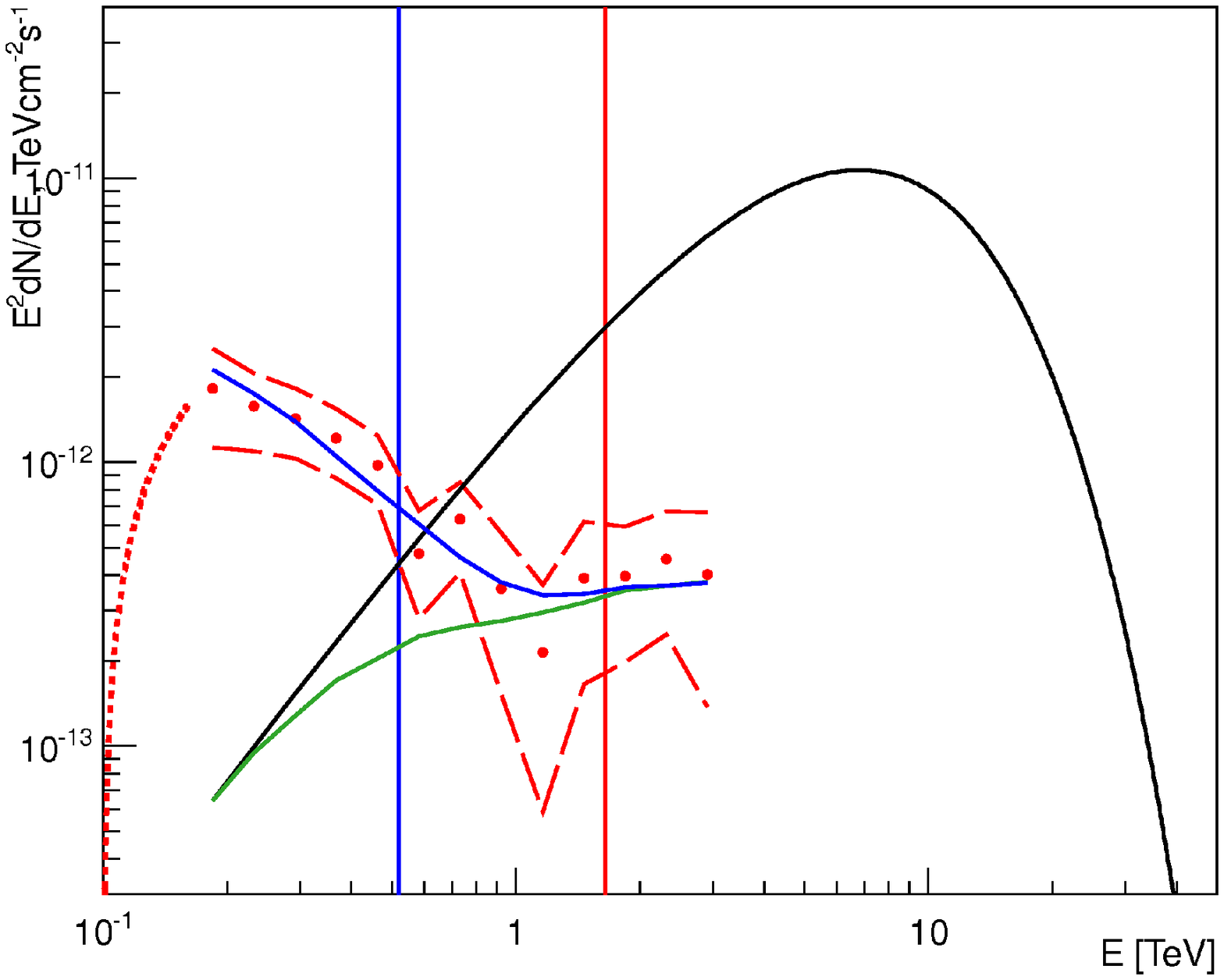}
\caption{\label{label} The fit to the 1ES 1101-232 SED including cascade component (blue curve). Figure from \cite{dzh15}.}
\end{minipage} 
\end{figure}

To explain the data in the framework of the CR beam model, an additional component from secondary photons is needed (the corresponding region of the spectrum is denoted by thick blue dashed lines). Now let us put a simple lower limit to the ratio of total power of accelerated CR $W_{CR}$ to the same quantity for VHE electrons $W_{VHE-e}= W_{VHE-\gamma}/f_{IC}$, where $f_{IC}$ is the fraction of energy of VHE electrons transferred to VHE photons by means of IC scattering. An important note is that gamma-rays produced by electrons deep inside blazar jets are expected to be highly beamed. On the other hand, a region of highly turbulent magnetic field (a ``lobe'') is expected to exist at the end of a jet, and CR particles are likely to be isotropized before they could produce secondary photons close to the Earth. Indeed, such ``lobes'' are observed in a number of radiogalaxies, and are expected to be able to scatter (or even confine) protons up to $10^{20}$ $eV$ \cite{hil84}. Therefore, if the observer is located near to the jet's axis, the observed flux from CR is effectively deamplified to a factor of $f_{beam}\sim 10^{3}$--$10^{4}$, assuming the angular radius of the jet $\theta_{Jet}$=1--3$^{\circ}$. Additionally, only a small fraction $f_{int}\sim 10^{-3}$ \cite{ess10} of CR energy is converted to photons near the observer.

Assuming the intrinsic VHE gamma-ray spectrum $dN/dE\sim E^{-2}$ (what is close to the shape of black curve in Figure 2) from 100 $GeV$ to 100 $TeV$, one may estimate that the secondary photons constitute a fraction $f_{sec} \sim10^{-2}$ of energy of intrinsic VHE photons. Therefore, assuming $f_{IC}= 0.1$, we find that $W_{CR}/W_{VHE-e}\sim f_{sec} \cdot f_{beam} \cdot (f_{IC}/f_{int})$= 10$^{3}$--10$^{4}$, depending of the $f_{beam}$ value, i.e. acceleration of hadrons is very effective, while for VHE electrons it is not. This challenges conventional models of gamma-ray blazars. Additionally, assuming $W_{VHE-e}= 10^{44}$--$10^{45} erg/s$, we find $W_{CR}$= 10$^{47}$-- 10$^{49} erg/s$, a very large value. The last values are similar to those obtained by \cite{raz12}.

\section{The EMC model}

\subsection{General considerations}

The primary photon absorption process $\gamma\gamma \rightarrow e^{+}e^{-}$ has threshold energy $E_{\gamma0}= m_{e}^{2}/\epsilon$, where $m_{e}$ is electron mass, and $\epsilon$ is EBL photon energy (we assume $c$=1 here). Most of pair-production acts occur not far from the threshold; in this case each produced electron receives energy $E_{e}\sim E_{\gamma0}/2$. Electrons, in their turn, produce secondary (cascade) photons by IC scattering. This process, in contrast to pair-production, doesn't have kinematical threshold, and occurs mainly on dense, low-energy cosmic microwave background (CMB) photon field; the energy of secondary photons $E_{\gamma-casc}\sim 3\cdot(E_{e}/1$ $TeV)^{2}$ $GeV$ \cite{kac12}. In this work we consider the case $100$ $GeV$ $<E_{\gamma0}<$ 100 $TeV$, thus $E_{\gamma-casc}<$ 7.5 $TeV$ $<< E_{\gamma0}$; for the case of $E_{\gamma0}= 10$ $TeV$ $E_{\gamma-casc}= 75$ $GeV$, again $<< E_{\gamma0}$. Therefore, the cascade process occurs in a peculiar, ``degenerate'' regime, and a typical number of cascade generations is small, $n_{Int}=$1--2.

In absence of EGMF the angular radius of cascade $\theta_{Int} \sim n_{Int}^{0.5}/\Gamma$, where $\Gamma= E_{e}/m_{e}> 10^{5}$ --- Lorentz factor of a cascade electron with $E_{e}>$100 $GeV$= 10$^{5}$ $MeV$. As we have seen, $\theta_{Jet} \sim 1^{\circ} \sim 10^{-2}$ rad; then $\theta_{Int}<10^{-5}<< \theta_{Jet}$, therefore the cascade process weakly modifies the angular distribution of photons from the source and 1D assumption that we shall use is well justified. Recently it was argued that plasma beam instabilities could dominate the energy losses for intergalactic cascade electrons \cite{bro12}. However, this mechanism is under debate: it was noted that the relativistic length contraction effect ``compresses the electric and associated magnetic fields'' around charged particles, and so, in fact, they do not constitute a plasma \cite{ven13}. For this reason, we do not include any electron energy loss process, except IC, in our calculations.

\subsection{The signatures of cascade emission}

Now let us illustrate the basic features of the EMC model of the VHE anomaly (figure 3). A fit to the measured SED was performed \cite{dzh15}, including the secondary component from electromagnetic cascades. The notations are the same as in figure 2, but the total model SED is now denoted by blue curve; the model SED without account of the cascade component is drawn by green curve. The impact of cascades to the SED is the difference between the blue and green curves. Blue vertical line denotes the ``gamma-ray horizon'' (the energy, where $\tau_{\gamma\gamma}$=1); red vertical line --- the energy, defined by condition $\tau_{\gamma\gamma}$=2. As in figure 2, the intrinsic SED in figure 3 was normalized to the absorbed one at $E$= 200 $GeV$.

The above-discussed rule $E_{\gamma-casc}<<E_{\gamma0}$ is clearly visible in figure 3, and the cascade component contributes mainly in the optically thin regime. However, to achieve a good fit, we were compelled to take a very hard primary spectrum, much harder, than in figure 2 for the same object neglecting the secondary component. Therefore, the model intensity in the optically thick regime in figure 3 (with account of the cascade component) is much higher, than in figure 2 (without this component), and for $\tau_{\gamma\gamma}>$2 the fit is in much better agreement with observations. It is quite remarkable that an exceptionally good fit of the shape of the spectrum in figure 3 (13 energy bins) was achieved with a very simple parametrization of the primary spectrum $dN/dE\propto E_{0}^{-\gamma}\cdot exp(-E_0 /E_c)$ with only two parameters $\gamma$ and $E_c$. As well, the significance of the anomaly for $\tau_{\gamma\gamma}>$2 drops from 2.1 $\sigma$ for the fit shown in figure 2 (without secondary component) to only 0.5 $\sigma$ for the new fit that includes the cascade component (for the significance evaluation method see \cite{dzh15}).

A possible influence of EGMF to the observed spectrum is schematically shown in Figure 3 by red dashed curve. In order to not spoil the fit, the EGMF strength $B$ should be sufficiently small, below some value $B_{Max}$. Recently A. Neronov et al. \cite{ner12} performed calculations of this ``magnetic cutoff'' shape for the case of blazar Mkn 501 ($z$= 0.034), and found that for $B=10^{-16}$ $G$ the cutoff starts to be detectable below $E\sim100$ $GeV$. Given that, as was noted above, the number of generations in cascade is small, and that the last low-energy point of the observed spectrum, located at $E\sim 200$ $GeV$, doesn't show a prominent magnetic cutoff, we can estimate $B_{Max} \sim (1-2)\cdot 10^{-16}$ $G$. Plausibility of such $B$ values will be discussed below in section 4. 

A spectrum of blazar in the considered version of the EMC model has three characteristic features, or signatures: a) low-energy magnetic cutoff b) the ``dip'', or a kind of an ``ankle'' that is usually located in the energy region where  $1<\tau_{\gamma\gamma}<2$ c) high-energy cutoff (not shown in figure 3). The third feature, the high-energy cutoff, was already studied in \cite{mur12}, \cite{tak13}. The detailed properties of this signature are not known, but the general conclusion that, in principle, allows to discriminate between the CR beam model and the EMC model, is the following: for large $\tau_{\gamma\gamma}$ (say, $>$2-3) the high-energy cutoff is much more marked for the case of EMC model \cite{mur12}, \cite{tak13}. The dip or ankle feature, as well, exists in both the CR beam (see, e.g., \cite{ess11}) and the EMC model. However, the low-energy magnetic cutoff is much more marked in the EMC model when compared to the CR beam model, unless the secondary component produced by CR dominates the entire VHE spectrum. We will see in section 4 that existing hints at intergalactic cascades tend to favour the EMC model. 

\section{Some constraints on the EGMF strength}

As we have seen, by now there exist only two conservarive models that could relax the VHE anomaly in blazar spectra, not to mention the option that the EBL intensity models must be revised: the CR beam model and the EMC model. Both require sufficiently low EGMF strength in voids to be plausible: $B<10^{-14}$ $G$ for the former \cite{pro12}, and even lower values (usually around $B<10^{-16}$ $G$) for the latter.

Any conclusive measurement of EGMF strength is still absent; some constraints on this quantity are presented in figure 4 (usually they are obtained for the EGMF coherence length 1 $Mpc$ \cite{aka10}). Upper bounds on the $B$ value are scarce and rather weak; \cite{bla99} obtained a constraint $B<10^{-9}$ $G$. $B= 2\cdot10^{-12}$ $G$ was found to be sufficient to explain magnetic fields in galaxy clusters \cite{dol05}. As well, we present a number of results obtained from non-observation of cascade component in blazar spectra at comparatively low energies, $E<$100 $GeV$. These constraints are highly model-dependent; \cite{arl14} even found that the zero EGMF hypothesis cannot be firmly rejected. Anyway, we present the cascade constraints obtained by \cite{der11}--\cite{vov12} in figure 4; these lower limits are in the range $B>10^{-18}$--$10^{-17}$ $G$. By analysing the angular distribution of arriving photons, \cite{abr14} performed a search of a magnetically broadened cascade pattern and was able to exclude the range of $B=3\cdot10^{-16}$--$10^{-14}$ $G$ at 99 \% C.L. Finally, by thick red horizontal line we show the $B=10^{-14}$ $G$ value, above which both the CR beam model and the EMC model are ruled out \cite{pro12}.  Future instruments with high sensitivity (e.g. CTA \cite{ach13}), or with good angular resolution, such as emulsion gamma-ray telescope \cite{aok12} or the GAMMA-400 detector \cite{gal13}, would be able to employ the same method as \cite{abr14} to put more tight constraints on the EGMF strength. Other EGMF searches based on stacking analysis of sources' angular distribution \cite{che14} or diffuse gamma-ray sky studies \cite{tas14}--\cite{che14b} do exist.

\begin{figure}[h]
\includegraphics[width=17pc]{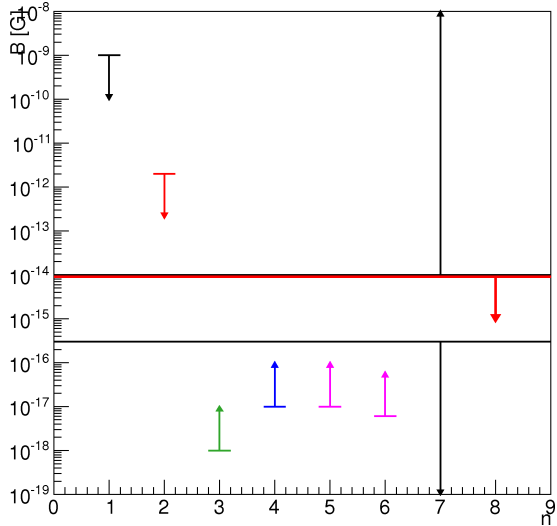}\hspace{2pc}%
\begin{minipage}[b]{19pc}\caption{\label{label} Some constraints on the EGMF strength in voids. $n$ denotes the number of corresponding work.
$n$=1 --- \cite{bla99}; $n$=2 --- \cite{dol05}; $n$=3 --- \cite{der11}; $n$=4 --- \cite{tay11}; $n$=5 --- \cite{vov12}, the case of the EBL model \cite{fra08}; $n$=6 --- \cite{vov12}, the case of the EBL lower limit from direct source counts; $n$=7 --- \cite{abr14} $n$=8 --- \cite{pro12}.}
\end{minipage}
\end{figure}

Finally, let us mention some recent results that actually give hints at the EM cascade presence in blazar spectra. Analysing the Mkn 501 spectrum, \cite{ner12} found that it could be explained by the intergalactic cascade hypothesis with $B=10^{-17}$--$10^{-16}$ $G$. Very recently, it was shown that gamma-ray sources with hard spectra are predominantly located in directions to voids \cite{fur14}. Moreover, it was found by \cite{fur14} that the EBL intensity fluctuations are not sufficient to explain the observed effect. Therefore, as first suggested in \cite{dzh15}, the most likely explanation is that \cite{fur14} actually observed the low-energy magnetic cutoff of secondary component, produced by EM cascade. This feature of the EMC model is schematically shown by dashed red line in figure 3. For the case of other objects, that do not point to voids, the mean $B$ value is likely much greater, and the cascade component may be greatly suppressed. For this reason, these latter objects do not show the spectral hardening observed by \cite{fur14}.

In principle, two other explanations of this effect are possible. The first is that the electromagnetic cascade upscatters the secondary photons to high enough energy to produce observable hardening. However, this is highly unlikely, given that a typical energy of secondary photon is much smaller than of the primary one, as discussed in subsection 3.1. The other explanation, that the effect observed by \cite{fur14} is caused by the CR-initiated cascade, is unlikely as well, unless the secondary component from these cascades dominates the VHE spectrum.

\section{Conclusions}

In the present work a brief review of conservative models that could explain the apparent faintness of the pair-production features in blazar spectra, was given. No existing model was found to be completely free of difficulties. The source-intrinsic models, while they are able to form a very hard intrinsic spectrum, do not explain other features of the anomaly (subsection 2.2). The model with production of secondaries by CR near to the Earth seems to require either a very high $W_{CR}/W_{VHE-e}$ ratio, or extremely good collimation of accelerated nuclei (subsection 2.3). Finally, the EM cascade (EMC) model was considered (section 3), that naturally explains some spectra with only two free parameters, doesn't contradict to contemporary constraints on the EGMF strength, and predicts the magnetic cutoff feature that may be already observed \cite{fur14}. Future observations will help to test this model.

\subsection*{Acknowledgments}
The work was supported by the Russian President grant LSS-3110.2014.2. The author is grateful to Prof. A. Kusenko for helpful discussions, to the anonymous referee for comments, and to Dr. V.V. Kalegaev, V.O. Barinova, M.D. Nguen, D.A. Parunakyan for permission to use the SINP MSU space monitoring data center computer cluster.


\section*{References}

\begin{thebibliography}{999}
\bibitem{tev15}
http://tevcat.uchicago.edu/
\bibitem{ack11}
Ackermann M et al. (Fermi LAT) 2011 {\it ApJ} {\bf 743} 171
\bibitem{nik62}
Nikishov A I 1962 {\it Sov. Phys. JETP} {\bf 14} 393
\bibitem{gou67}
Gould R J and Shreder G 1967 {\it Phys. Rev.} {\bf 155} 1408
\bibitem{ack12}
Ackermann M et al. (Fermi LAT) 2012 {\it Science} {\bf 338} 1190
\bibitem{abr13}
Abramowski A et al. (H.E.S.S.) 2013 {\it A\&A} {\bf 550} A4
\bibitem{hor12}
Horns D and Meyer M 2012 {\it JCAP} 02 033
\bibitem{aha99}
Aharonian F A et al. (HEGRA) 1999 {\it A\&A} {\bf 349} 11
\bibitem{pro00}
Protheroe R J and Meyer H 2000 {\it Phys. Lett.} B {\bf 493} 1
\bibitem{kif99}
Kifune T 1999 {\it ApJ} {\bf 518} L21 
\bibitem{ame00}
Amelino-Camelia G and Piran T 2001 {\it Phys.Lett.} B {\bf 497} 265
\bibitem{sik83}
Sikivie P 1983 {\it Phys. Rev. Lett.} {\bf 51} 1415 
\bibitem{mir07}
Mirizzi A et al. 2007 {\it Phys. Rev.} D {\bf 76} 023001
\bibitem{san09}
Sanchez-Conde M A et al. 2009 {\it Phys. Rev.} D {\bf 79} 123511
\bibitem{dzh15}
Dzhatdoev T 2015 {\it Izvestiya Rossiiskoi Akademii Nauk} {\bf 79} 363 (In print) ({\it Preprint} astro-ph/1501.00259)
\bibitem{hor13}
Meyer M et al. 2012 Revisiting the Indication for a low opacity Universe for very high energy gamma-rays {\it Preprint} astro-ph/1211.6405
\bibitem{kne10}
Kneiske T M and Dole H 2010 {\it A\&A} {\bf 515} A19
\bibitem{kne04}
Kneiske T M et al. 2004 {\it A\&A} {\bf 413} 807
\bibitem{ste06}
Stecker F W et al. 2006 {\it ApJ} {\bf 648} 774
\bibitem{fra08}
Franceschini A et al. 2008 {\it A\&A} {\bf 487} 837
\bibitem{pri08}
Primack J R et al. 2008  {\it AIP Conference Proceedings} {\bf 1085} 71
\bibitem{dom10}
Dominguez A et al. 2011 {\it MNRAS} {\bf 410} 2556
\bibitem{fin10}
Finke J D et al. 2010 {\it ApJ} {\bf 712} 238
\bibitem{gil12}
Gilmore R C et al. 2012 {\it MNRAS} {\bf 422} 3189
\bibitem{aha02}
Aharonian F A et al. 2002 {\it A\&A} {\bf 384} 834
\bibitem{aha08}
Aharonian F A et al. 2008 {\it MNRAS} {\bf 387} 1206
\bibitem{oik14}
Oikonomou F et al. 2014 {\it A\&A} {\bf 568} A110
\bibitem{pad97}
Padovani P 1997 {\it Very High Energy Phenomena in the Universe; Morion Workshop. Eds. Y. Giraud-Heraud and J. Tran Thanh Van, 1997} 7
\bibitem{wou14}
Wouters D and Brun P 2014 {\it JCAP} 01 016
\bibitem{ury98}
Uryson A 1998 {\it JETP} {\bf 86} 213
\bibitem{ess10}
Essey W and Kusenko A 2010 {\it APh} {\bf 33} 81
\bibitem{ave07}
d'Avezac P et al. 2007 {\it A\&A} {\bf 469} 857
\bibitem{ess10b}
Essey W et al. 2010 {\it Phys. Rev. Lett.} {\bf 104} 141102
\bibitem{ess11}
Essey W et al. 2011 {\it ApJ} {\bf 731} 51
\bibitem{mur12}
Murase K et al. 2012 {\it ApJ} {\bf 749} 63
\bibitem{ess14}
Essey W and Kusenko A 2014 {\it APh} {\bf 57} 30 
\bibitem{aha13}
Aharonian F A et al. 2013 {\it Phys. Rev.} D {\bf 87} 063002
\bibitem{pro12}
Prosekin A et al. 2012 {\it ApJ} {\bf 757} 183 
\bibitem{kal13}
Kalashev O E et al. 2013 {\it Phys. Rev. Lett.} {\bf 111} 041103
\bibitem{kac12}
Kachelriess M et al. 2012 {\it Comp. Phys. Comm.} {\bf 183} 1036
\bibitem{aha06}
Aharonian F A et al. (H. E. S. S.) 2006 {\it Nature} {\bf 440} 1018
\bibitem{hil84}
Hillas A M 1984 {\it ARA\&A} {\bf 22} 425
\bibitem{raz12}
Razzaque S et al. 2012 {\it ApJ} {\bf 745} 196
\bibitem{bro12}
Broderick A E et al. {\it ApJ} {\bf 752} 22
\bibitem{ven13}
Venters T M and Pavlidou V 2013 {\it MNRAS} {\bf 432} 3485
\bibitem{ner12}
Neronov A et al. 2012 {\it A\&A} {\bf 541} A31
\bibitem{tak13}
Takami H et al. 2013 {\it ApJ Lett.} {\bf 771} L32
\bibitem{aka10}
Akahori T and Ryu D 2010 {\it ApJ} {\bf 723} 476
\bibitem{bla99}
Blasi P et al. 1999 {\it ApJ} {\bf 514} L79
\bibitem{dol05}
Dolag K et al. 2005 {\it JCAP} 01 009
\bibitem{arl14}
Arlen T C et al. 2014 {\it ApJ} {\bf 796} 18
\bibitem{der11}
Dermer C D et al. 2011 {\it ApJ Lett.} {\bf 733} L21
\bibitem{tay11}
Taylor A M et al. 2011 {\it A\&A} {\bf 529} A144
\bibitem{vov12}
Vovk Ie et al. 2012 {\it ApJ Lett.} {\bf 747} L14
\bibitem{abr14}
Abramowski A et al. (H.E.S.S.) 2014 {\it A\&A} {\bf 562} A145
\bibitem{ach13}
Acharya B S et al. (CTA) 2013 {\it APh} {\bf 43} 3
\bibitem{aok12}
Aoki S et al. 2012 Balloon-borne gamma-ray telescope with nuclear emulsion: overview and status {\it Preprint} astro-ph/1202.2529
\bibitem{gal13}
Galper A M et al. (GAMMA-400) 2013 {\it AIPCP} {\bf 1516} 288
\bibitem{che14}
Chen W et al. 2014 Evidence for GeV Pair Halos around Low Redshift Blazars {\it Preprint} astro-ph/1410.7717
\bibitem{tas14}
Tashiro H et al. 2014 {\it MNRAS Lett.} {\bf 445} L41
\bibitem{che14b}
Chen W et al. 2014 Intergalactic magnetic field spectra from diffuse gamma rays {\it Preprint} astro-ph/1412.3171
\bibitem{fur14}
Furniss A et al. 2015 {\it MNRAS} {\bf 446} 2267
\end{thebibliography}
\end{document}